\documentclass[12pt]{article}

\catcode`\@=11

\global\arraycolsep=2pt
\usepackage{amsbsy,amssymb,latexsym,amsfonts,amsmath}
\usepackage{graphicx,color}

\begin{document}
\begin{flushright}
\parbox{4.2cm}
{CALT-TH 2015-031}
\end{flushright}

\vspace*{0.7cm}

\begin{center}
{\Large Strong gauging or decoupling ghost matter}
\vspace*{1.5cm}\\
{Yu Nakayama}
\end{center}
\vspace*{1.0cm}
\begin{center}
{\it Walter Burke Institute for Theoretical Physics, California Institute of Technology,  \\ 
Pasadena, California 91125, USA}
\vspace{3.8cm}
\end{center}

\begin{abstract}
Gauging extra matter is a common way to couple two CFTs discontinuously. We may consider gauging matter by strongly coupled gauge theories at criticality
rather than by weakly coupled (asymptotic free) gauge theories.
It often triggers relevant deformations and possibly leads to a non-trivial fixed point. In many examples such as the IR limit of SQCDs (and their variants), the relevant RG flow induced by this strong gauging makes the total central charge $a$ increase rather than decrease compared with the sum of the original decoupled CFTs. The dilaton effective field theory argument given by Komargodski and Schwimmer does not apply because  strong gauging is not a simple deformation by operators in the original two decoupled CFTs and it may not be UV complete. When the added matter is vector-like, one may emulate strong gauging in a UV completed manner by decoupling of ghost matter. While the UV completed description makes the dilaton effective field theory argument possible, due to the non-unitarity, we cannot conclude the positivity of the central charge difference in accordance with the observations in various examples that show the contrary.
\end{abstract}

\thispagestyle{empty} 

\setcounter{page}{0}

\newpage

\section{Introduction}
A gauge symmetry plays a central role in modern physics. 
Within a Lagrangian description, this fictitious ``symmetry" turns out to be imperative in order to describe massless particles with spin $\ge 1$. In $d=4$ dimensions, all the perturbatively renormalizable unitary Lagrangian field theories without gauge interaction are believed to flow down to trivial non-interacting fixed points in the infrared (IR), so non-trivial critical phenomena in $d=4$ dimensions should rely on gauge symmetries within a Lagrangian description.

Gauging a global symmetry is a common  way to construct new quantum field theories, and it has been heavily used to explore non-trivial constructions of (super)conformal fixed points and various dualities among them. Typically, the added gauge interaction is assumed to be asymptotic free and weakly coupled in the ultra-violet (UV) limit. Then we have a Lagrangian description of gauging (to be called ``weak gauging" in this paper) schematically described by adding the gauging interaction $\int d^4x A_\mu J^\mu + \cdots$ in the weakly coupled asymptotic free  UV limit of the gauge coupling. Note that the ``gauged matter" (represented by the conserved current operator $J^\mu$ before gauging) does not have to be weakly coupled by itself. As long as matter has a global symmetry and the gauging symmetry is anomaly free, weak gauging should make sense.\footnote{Strictly speaking, to our knowledge we do not know the formal proof of this Noether procedure in particular in non Lagrangian theories since we may encounter potential higher order obstructions. Gauging a massless spin $3/2$ field may be one example of such rare occasions where inconsistency may happen.}
 The subsequent renormalization group (RG) flow makes the gauge coupling constant larger toward the IR and it may hit a new RG fixed point realized by a conformal field theory (CFT).\footnote{In this paper, we assume that the RG fixed point shows conformal invariance. See e.g. \cite{Nakayama:2013is} on this point.} 
One may study the RG flow by various non-perturbative techniques available.

While this weak gauging is a UV complete procedure, it is not necessarily what we might want to follow. 
Suppose that the original ``gauging theory" ($A$) is given by a gauge theory coupled with some matter, which has already flowed down to a non-trivial strongly coupled IR fixed point ($A'$). We would like to couple another (potentially strongly coupled) CFT ($B$) with a global symmetry by gauging this symmetry from gauge fields in $(A')$. This gauging may subsequently trigger relevant deformations and possibly lead to a non-trivial fixed point $(C)$.
Of course, we could always do it if we go back to the UV limit $(A)$ with a Lagrangian description and couple matter CFT ($B$) by the method of weak gauging as described in the previous paragraph.
However this is not always satisfactory because (1) we have to rely on the Lagrangian description of UV theory ($A$),  
(2) we do not use non-trivial information contained in the fixed point ($A'$), and (3) the induced RG flow may not be economical and it may be hard to follow. 

Let us compare them with the situations in the Landau-Ginzburg flow in $d=2$ dimensions. An analogue of theory ($A$) is the UV description as a perturbation of a free massless scalar $\phi$ by the potential $V(\phi)=\phi^{2p} + \cdots$ (with lower terms fine-tuned).
It flows to a  $(p+1,p+2)$ minimal model in the IR, which is an analogue of theory ($A'$). 
We want to study the further flow of this theory by changing $p$ by one unit. The UV completed description similar to the above weak gauging is to go back to the UV Lagrangian description and add $\tilde{V}(\phi) = \phi^{2p-2}$ (with lower terms fine-tuned) to study the flow from the beginning again. However, it is much more economical to study the conformal perturbation theory directly around $(A')$ by  the $O_{1,3}$ operator of the $(p+1,p+2)$ minimal model. 
For large $p$, these two fixed points are very closely located and the conformal perturbation theory is much easier to follow than the UV Lagrangian flow. Even in the strongly coupled case for smaller $p$, the truncated conformal space method will give a much better way to study the RG flow than the UV Lagrangian description based on the free massless scalar. Actually, this direct flow is even integrable while the UV flow from the free massless scalar theory is not.

The question arises whether we can offer a similar analysis in gauging extra matter. 
Take the Banks-Zaks theory for example. We start with the IR fixed point of the $SU(N_c)$ gauge theory with $N_f$ massless Dirac fermions in the fundamental representation (typically called ``quarks"). Can we add another massless quark? The additional gauged quark will lead to a new fixed point with $N_f+1$ flavors. Weak gauging perspective is to go back to the UV theory and then add the new quark and study the flow from the beginning. 
As is the case with the minimal model flow, this does not sound very economical. Can we bypass the procedure and directly gauge the extra flavor at the non-trivial Banks-Zaks fixed point?  Can we directly study the RG flow induced by gauging in the language of the original decoupled two CFTs without referring to the UV Lagrangian? After all, the RG flow between two Banks-Zaks fixed points with different number of flavors may be much shorter than going back to the UV fixed point and then flowing down. It would be wonderful if sufficient knowledge of the fixed point before gauging gives some insight of what will happen after gauging as in the $O_{1,3}$ integrable deformation in the Landau-Ginzburg flow.

In this paper, we name gauging by a strongly coupled gauge theory at criticality as ``strong gauging", and pursue such possibilities. One motivation is to explore how much known methods (e.g. conformal perturbation theory, truncated conformal methods, duality, conformal bootstrap, supersymmetric methods) that do not rely on the UV Lagrangian can be applied to general gauging procedures in principle. We are also interested in the subsequent RG flow e.g., if it satisfies generally expected properties such as the $a$-theorem or the gradient flow.

In section 2, we will propose several ways to implement strong gauging and discuss pros and cons. In section 3, we study the properties of the subsequent RG flow after strong gauging. In particular we  discuss the validity of the $a$-theorem. In section 4, we present various examples of strong gauging realized in supersymmetric gauge theories in which non-perturbative computations of the central charges are possible. In section 5, we present a novel way to emulate strong gauging by using a method of decoupling ghost matter. This provides a UV completed understanding of strong gauging.

\section{Implementation of strong gauging}

In this section, we would like to discuss the implementation of strong gauging in quantum field theories and its possible obstructions.
Before going into the actual implementation, however, we should start with what we mean when we say gauging by strongly coupled gauge theories. 
In particular, if we try to implement gauging by field theories at criticality, the definition may not be unique.

A potential problem is as follows: from the abstract local CFT data (e.g. spectrum and OPE coefficients) of theory $(A')$ without referring to UV theory $(A)$, how do we know whether theory $(A')$ contains the gauge field we can use  and which symmetries we can gauge? 
Without a Lagrangian description, how do we know that this theory can gauge only $SU(5)$ rather than $SU(2)$ global symmetries?
Actually, the situation is more non-trivial because which gauge symmetry this IR theory $(A')$ can gauge may depend on the UV completion. Take the non-trivial IR fixed point of SQCD for instance. We may realize it as an IR limit of an asymptotic free electric theory with the $SU(N_c)$ gauge group (with fundamental quark chiral superfields) or an IR limit of the asymptotic free magnetic theory of $SU(N_f-N_c)$ gauge group (with fundamental quark chiral superfields and dual meson superfields). Depending on the realization we choose, the possibility of gauging seems different.

We do not regard it as a serious issue. Rather we take it as an advantage. Suppose we have multiple ways to realize the same CFT from different gauge groups.  If we would like to perform strong gauging by this CFT, we will just pick up one realization, and add extra matter in accordance with the chosen gauge symmetry.  More possibilities to realize the same CFT just gives us more ways to introduce different matter in oder to obtain new critical phenomena. For this purpose, it is clearly not a disadvantage, rather it is an advantage to have many different realizations and more possible new CFTs after gauging extra matter.

We still have to discuss the following point, however.
Even if we choose a particular gauge theory realization of the CFT, by definition this information is not encoded in the local CFT data (otherwise the non-uniqueness problem has been already solved).
Accordingly the added interaction in gauging is not a well-defined operator within the two original CFTs $(A')$ and $(B)$. Concretely, in strong gauging, we would like to add the deformation $\int d^4x A_\mu J^\mu + \cdots$ to couple the two CFTs, but the ``gauge potential" $A_\mu$ is not a well-defined operator in theory $(A')$ viewed as a local CFT. In addition, there may be more gauge invariant operators like $O_{(A')} O_{(B)}$, where $O_{(A')}$ and $O_{(B)}$ are not gauge invariant separately. Depending on the other physical requirement (such as requirement from the extra symmetry or as a local counter-term) we may also add them as a part of the definition of strong gauging.

One possible procedure goes as follows. We first pick up a particular gauge theory realization of the CFT data. What this really means is that we extend the Hilbert space (e.g. operator algebra in the radial quantization) including unphysical states (e.g. ghost sector) with the BRST charge so that the BRST cohomology corresponds to the original CFT states. When the fixed points are perturbative (such as the Banks-Zaks fixed point \cite{Caswell:1974gg}\cite{Banks:1981nn}), this procedure is not difficult because the extended Hilbert space are continuously connected to the perturbative quantum field theories which are well known in the BRST quantization of gauge theories (with matter).
Indeed, this BRST quantization is nothing but how we compute the RG beta functions and the location of the fixed points by using the perturbation theory  in asymptotic free theories. 
Recall that this BRST extension is not unique; it depends on the gauge fixing and the definition of the BRST charge, but the BRST cohomology must be same. 
We also note that the extended Hilbert states may not possess conformal invariance even though the original fixed point is conformally invariant (see e.g. \cite{ElShowk:2011gz}) because the gauge fixing term does not necessarily preserve conformal invariance. In principle, we may be able to compute all the correlation functions in this extended Hilbert states such that the BRST cohomology encodes the CFT data.
Now in this extended Hilbert space, the operator $\int d^4x A_\mu J^\mu + \cdots$ that is used in strong gauging makes sense, and we may use the information contained in the extended Hilbert space to study the RG flow after strong gauging. 
In Lagrangian gauge theories, this is just the standard way of gauging matter, and there is no much difference compared with weak gauging in asymptotic free gauge theories.

In this viewpoint, the idea of gauging matter by strongly coupled gauge theories is nothing new. Furthermore, if we consider gauge theories in $d>4$ dimensions (e.g. \cite{Intriligator:1997pq}) strong gauging is unavoidable.  In particular, in lattice gauge theories, the strongly coupled expansion has been discussed since its birth  \cite{Wilson:1974sk} (see also \cite{Drouffe:1983fv} for a review). They have a fixed UV regularization (by lattice) and they introduce the gauge fields and matter at each lattice site (or link) in the strongly coupled limit of the bare coupling constant.
The path integral over the lattice  with the strongly coupled gauge kinetic term is well-defined.\footnote{In lattice regularization, we do not necessarily have to fix the gauge because the gauge volume is finite, but we may fix the gauge and perform the BRST quantization if we wish.}  Actually, it is even easier to do the path integral because we can employ the high temperature expansions of lattice systems

One concrete realization of strong gauging on the lattice therefore goes like this. We begin with the UV lattice with gauge fields on the links and matter on the sites. We follow the RG flow to reach the non-trivial fixed points below a certain scale. Now, we take the sublattice of the original lattice and put extra matter only on the sublattice. Below the scale of the sublattice, the gauge fields feel the presence of extra matter, and  we realize strong gauging. The UV RG flow is not changed by extra matter on the sublattice because it does not cause any UV fluctuations.

There is a complication we should point out, however. This definition does not necessarily gives the continuum limit of strong gauging, and the physics we would get in the low energy limit may not be universal. 
Therefore, the insight we would obtain in the strongly coupled limit of the lattice regularization could be remote from what we would expect in confining gauge theories such as QCD. This is because we are more interested in the non-universal features such as the hadron masses that are UV sensitive unless we take the continuum limit. Therefore it is crucial to take the continuum limit in the lattice QCD simulations.

Things are better tamed if the gauged theory possesses a non-trivial IR fixed point. Due to the universality, as long as we are interested in the local CFT data of the IR fixed point, we do not have to take the continuum limit. The long wavelength limit automatically picks out the universal feature of the system. The implementation of strong gauging by extending the Hilbert space and adding the non BRST invariant operator of the original two decoupled systems may not be UV complete. However, if the resulting RG flow leads to a non-trivial IR fixed point and we are only interested in the universal aspects of the new IR fixed points, it is not a problem. We are only interested in such situations in this paper.

The same discussions apply when we would like to realize gauge theories in the truncated conformal space approach on the cylinder $\mathbf{S}^d \times \mathbf{R}$ (see \cite{Hogervorst:2014rta} for a recent discussion). Although the truncated conformal space approach does not rely on the Lagrangian description in principle, the realization of gauge theories may require the introduction of the extended Hilbert space with unphysical states. Since the gauge fixing may break conformal invariance, we have to abandon the manifest conformal structure on the cylinder spectrum (see, however, \cite{Rychkov:2014eea} for a cylinder truncation approach without conformal symmetry). Furthermore, the BRST symmetry is typically broken due to the truncation.
This applies not only to the realization of strong gauging in the truncated conformal space approach, but also to weak gauging.\footnote{We would like to thank S.~Rychkov for the related discussions.}

Thus, while the truncated conformal space approach is an alternative to the lattice regularization, the gauge theories still require special treatment in comparison with the more conventional deformed conformal field theories such as the Landau-Ginzburg-Wilson fixed point. There, all the states are physically realized on $\mathbf{S}^d \times \mathbf{R}$ and the usefulness of the truncated conformal space approach was demonstrated in \cite{Hogervorst:2014rta}.
To aim at a possible improvement of the situation, we will propose an alternative way to emulate strong gauging by a method of decoupling ghost matter in section 5. The method may be applicable to the truncated conformal space approach since it preserves conformal invariance unlike the BRST gauge fixing. Moreover,  we may avoid careful treatment of the BRST symmetry, which can be broken by the truncation.
 Yet, we find that we cannot avoid the extension of the Hilbert states with a non-positive definite norm.




\section{Renormalization group flow from strong gauging and $a$-theorem}
Once we have implemented the idea of strong gauging, we would like to study the properties of the RG flow after gauging. The first question we ask is whether this strong gauging is a relevant deformation and whether it leads to a (possibly non-trivial) new fixed point. 
In the picture of weak gauging, as long as the gauged matter content is in the conformal window, the gauged theory flows down to a non-trivial IR fixed point. It is therefore expected that strong gauging will lead to the same IR fixed point.\footnote{We are always implicit about tuning the other coupling constants if necessary.}
In many examples that we will study, this is the case, but we should keep in mind that the set of relevant deformations may differ in weak gauging by theory $(A)$ and  strong gauging by theory $(A')$. In such situations, the end point of the RG flow may be different. 

Intuitively, at the non-trivial fixed point ($A'$), the matter contribution and the gauge contribution to the RG beta functions of the gauge coupling constant balance. If we add extra matter $(B)$ by strong gauging, the RG beta function of the combined system $(A') + (B)$ will have more contributions from added matter and it becomes positive. 
Then, the coupling constant starts to run such that the gauge coupling becomes weaker toward the IR. The novel IR fixed point $(C)$, if any, should be located in the weaker gauge coupling region with more matter. Indeed, this is what can be inferred in the perturbative computations of RG beta functions such as near the perturbative Banks-Zaks fixed point. 
This RG flow is fundamentally different from the weak gauging RG flow from $(A) + (B)$ to $(C)$ because in the latter the RG beta function is negative in the asymptotic free UV limit and the gauge coupling constant becomes larger toward the fixed point in the IR limit. 
On the other hand, the RG flow induced by strong gauging approaches the same fixed point from above as long as this intuitive picture is valid. 


One important aspect of the RG flow in relativistic quantum field theories (in $d=4$ dimension) is the $a$-theorem \cite{Cardy:1988cwa}\cite{Komargodski:2011vj}. The claim is that if the RG flow induced by the relevant deformations that connect two conformally invariant fixed points, the difference of the central charge $a_{\mathrm{UV}} - a_{\mathrm{IR}}$ between the undeformed theory and the one in the IR limit must be positive. The central charge is defined such that at the conformal fixed point it is given by the coefficient in front of the Euler term in the conformal anomaly\footnote{Our normalization is such that $a=\frac{1}{360},\frac{11}{360},\frac{62}{360}$ for a free real scalar, a free Dirac fermion and a free vector field.}
\begin{align}
\langle T^{\mu}_{\ \mu} \rangle =  \frac{a}{(4\pi)^2} \mathrm{Euler} - \frac{c}{(4\pi)^2} \mathrm{Weyl}^2 \ .
\end{align}
The proof of this theorem requires the unitarity and the UV completion that guarantees the positivity of the two-two dilaton scattering amplitudes that may be interpreted as the difference of the central charges \cite{Komargodski:2011vj}.
In many (perturbative) examples, even the stronger claim of the gradient flow condition holds \cite{Osborn:1991gm}, in which the RG beta function can be derived from 
\begin{align}
\frac{d g^i}{d\log\Lambda} = \beta^i = g^{ij} \partial_j \tilde{a} \label{grad}
\end{align}
 with the potential $\tilde{a}$ is identified as the $a$-function $\tilde{a} = a$ at the fixed point. Here $g_{ij}$ is a metric on the coupling constant space.

It is easy to argue that the $a$-theorem is satisfied in weak gauging. The central charge $a$ therefore always satisfies the inequalities $a_{(A)} > a_{(A')}$ and $a_{(A)} +a_{(B)} > a_{(C)}$. In both cases, gauging matter can be regarded as UV complete unitary deformations of theory $(A)$ and $(B)$. The coupling constant (e.g. gauge coupling constant) in weak gauging  appears in kinetic terms of the asymptotic free UV gauge fields, and it can be made Weyl invariant by dressing it with the background dilaton. Furthermore, the dilaton decouples in the UV limit as well as in the IR limit (as long as the IR fixed point is conformal). Therefore the argument given in \cite{Komargodski:2011vj} (see also \cite{Komargodski:2011xv}\cite{Luty:2012ww} in relation to this point) works and the central charge $a$ always decreases from $(A)$ to $(A')$ or from $(A) + (B)$ to $(C)$.

However, if we examine strong gauging from $(A') + (B)$ to $(C)$, we find that in most of the examples (which will be explicitly shown in section 4), the central charge $a$ { increases} rather than decreases (i.e. $a_{(A')} + a_{(B)}< a_{(C)}$) even though we argue that strong gauging induces a relevant deformation.

We can easily convince ourselves that this happens in the following example. 
Take the Banks-Zaks fixed point of $N_f$ massless Dirac fermions in the fundamental representation as theory $(A')$ such that $N_f$ is close to the lower edge of the conformal window of the $SU(N_c)$ gauge theory.\footnote{We do not know the actual critical flavor number, but it is not important.} Since the Banks-Zaks fixed point is strongly coupled, the central charge satisfies $a_{(A)} > a_{(A')}$, where UV theory $(A)$ is just given by $N_c^2 -1$ massless free vector fields and  $N_fN_c$ free massless Dirac fermions. More explicitly, we have \cite{Jack:1990eb}
\begin{align}
 a_{(A)} &= \frac{1}{360}\left(62(N_c^2-1) + 11N_fN_c \right) \cr
a_{(A')} &= a_{(A)} - \frac{1}{360}\left(\frac{8}{5}(\frac{11}{2}N_c-N_f)^2 \right) + \mathcal{O}\left( (\frac{11}{2}N_f-N_f)^3 \right) \ , \label{central}
\end{align}
where the latter is valid only when it is near the upper edge of the conformal window so that $N_f \simeq \frac{11}{2} N_c$. There is no known exact formula beyond the perturbation theory (see, however, the supersymmetric case in the next section where the exact formula is available).

Now, we will add $N_f'$ fundamental  flavors and strongly gauge them. The $N_f'$ is chosen so that $N_f+N_f'$ is close to the upper edge of the conformal window of the $SU(N_c)$ gauge theories (i.e. $N_f+N_f' \simeq\frac{11}{2} N_c$) and the resulting IR fixed point is very weakly coupled. Trivially, we have the inequality $a_{(A)} + a_{(B)} > a_{(A')} + a_{(B)}$ where $a_{(B)} = \frac{11}{360} N_cN_f'$ is the central charge of $N_cN_f'$ free massless Dirac fermions. 
On the other hand, we know that after the RG flow induced by strong gauging, we would end up with the weakly coupled Banks-Zaks fixed point with $N_f + N_f'$ flavors. Since we have chosen $N_f'$ so that it is close to the upper edge of the conformal window, the approximate relation $a_{(C)} \simeq a_{(A)} + a_{(B)}$ holds (see \eqref{central}). From the above inequality, we immediately conclude that $a_{(C)} > a_{(A')} + a_{(B)}$. Thus in this case, the central charge $a$ increases under the RG flow induced by strong gauging.

We should, however, note that the reason why the central charge increases rather than decreases cannot be attributed solely to the fact that we approach the new fixed point above from the strongly coupled region rather than below from the weakly coupled region. Indeed, the perturbative computation tells that even above the fixed point $g_*$, we may perturbatively construct the $a$-function that gives us the gradient flow for the gauge coupling constant \cite{Jack:1990eb}.\footnote{Of course, the meaning of the $a$-function as the potential function for the gradient flow here should be taken with a grain of salt because unless we have another zero of the RG  beta functions in the strongly coupled regime, we cannot assign the physical meaning to the $a$-function.} 
Since the metric $g_{ij}$ on the coupling constant space that appears in \eqref{grad} is perturbatively positive there (irrespective of whether we are above or below the fixed point), the RG flow would have shown the decrease of this particular $a$-function (which is the potential of the gradient flow \cite{Osborn:1991gm}). 

The point is that the gradient flow computation perturbatively done does not necessarily capture the physics of the RG flow induced by strong gauging  here. First of all, we cannot identify the $a$-function for the gradient flow for the gauge coupling constant perturbatively computed around the asymptotic free fixed point as the $a$-function of our strong gauging even if we are in the weakly coupled regime. One of the underlying assumptions in the perturbative computation of the $a$-function is that the gauged matter are all treated equally, and more specifically in the above example, the $N_f + N_f'$ fermions have the same wavefunction renormalizaiton factors. However, since the energy-scale we gauge the $N_f$ flavors and $N_f'$ flavors are different here, we cannot use the same formula of the $a$-function to generate the gradient flow for the RG beta function of the gauge coupling constant.  Actually, the difference of the wavefunction renormalization factors of $N_f$ flavors and $N_f'$ flavors should be regarded as another running coupling constant, which will become identical in the new Banks-Zaks fixed point they eventually reach. This parameter is an irrelevant deformation near the new Banks-Zaks fixed point.
Whether the RG flow in this larger coupling space is a gradient flow or not is a completely different question, and needs more study.\footnote{The fact that this deformation is an irrelevant deformation near the fixed point makes it harder for the analysis. See e.g. \cite{Gukov:2015qea} for recent discussions on renormalization group flow with (dangerously) irrelevant deformations.}

In section 4, we collect more examples in which we can compute the difference of the central charges before and after gauging in a non-perturbative manner. In section 5, we try to reconcile with what we know in the proof of the $a$-theorem and what we observe in strong gauging by proposing a way to emulate strong gauging by decoupling of ghost matter. There the decoupling ghost may accompany the gradient flow of the ghost mass term and the gauge coupling constant but the metric for the gradient flow will not be positive definite so that it may give an explanation why strong gauging avoids the $a$-theorem.

\section{More supersymmetric examples}

In this section, we will present  various examples where we can show that the central charge $a$ increases from the ``relevant" deformations of two CFTs by strong gauging.
Our focus is on the supersymmetric gauge theories in which one can compute the central charge $a$ before and after strong gauging non-perturbatively.

As the first example, let us take conformal SQCD with $SU(N_c)$ gauge group with $N_f$ fundamental flavors. 
We also add $2N_c \times N_f'$   free chiral ``spectator" superfields that we will eventually gauge.
For simplicity, we take the Veneziano limit, i.e.  $N_f, N_f', N_c \gg 1$ with the fixed ratio
\begin{align}
x &= \frac{N_f}{N_c} \cr
z& = \frac{N_f+N_f'}{N_c} \ .
\end{align}
The following analysis can be made without taking the Veneziano limit and this is just for simplicity of the presentation.
In order to obtain a non-trivial IR fixed point as theory $(A')$, we first assume $\frac{3}{2}<x<3$ and $x<z<3$. Later, we will discuss the situation in which the IR theory is in the magnetic IR free regime $1<x<\frac{3}{2}$.

The theory is asymptotic free, and UV free fixed point $(A)$  has the central charge \cite{Anselmi:1997am}
\begin{align}
a_{(A)} = \frac{N_c^2}{48}( 9 + 2x) \ .
\end{align}
up to $1/N_c$ corrections neglected in the Veneziano limit.
The IR fixed point $(A') + (B)$ before gauging is just a sum of conformal $SU(N_c)$ SQCD with $N_f$ flavors  $(A')$ and $2N_c \times N_f'$ free chiral multiplets $(B)$. 
The central charge can be computed \cite{Anselmi:1997am} as
\begin{align}
a_{(A')} + a_{(B)} &=  \frac{N_c^2}{48}(18 - 27x^{-2}  + 2(z-x)) 
\  \label{aA'}
\end{align}
up to $1/N_c$ corrections.
Obviously, $a_{(A)}>a_{(A')}$ and the $a$-theorem holds in this RG flow. We naturally regard weak gauging of $N_f$ fundamentals as a relevant deformation of the asymptotic free UV fixed point.

As an example of strong gauging, suppose we are at this non-trivial IR fixed point, and then try to add new flavors by gauging $2N_c \times N_f'$ free chiral multiplets $(B)$ that have been so far spectators. As we will argue below, this is  a relevant deformation and eventually leads to another IR fixed point $(C)$ given by conformal $SU(N_c)$ SQCD with $N_f + N_f'$ fundamental flavors. The central charge of the new fixed point can be computed as \cite{Anselmi:1997am} 
\begin{align}
a_{{(C)}} &= \frac{N_c^2}{48}(18 - 27z^{-2})  
 \label{aC}
\end{align}
up to $1/N_c$ corrections.

\begin{figure}[tbh]
\begin{center}
\includegraphics[width=\linewidth]{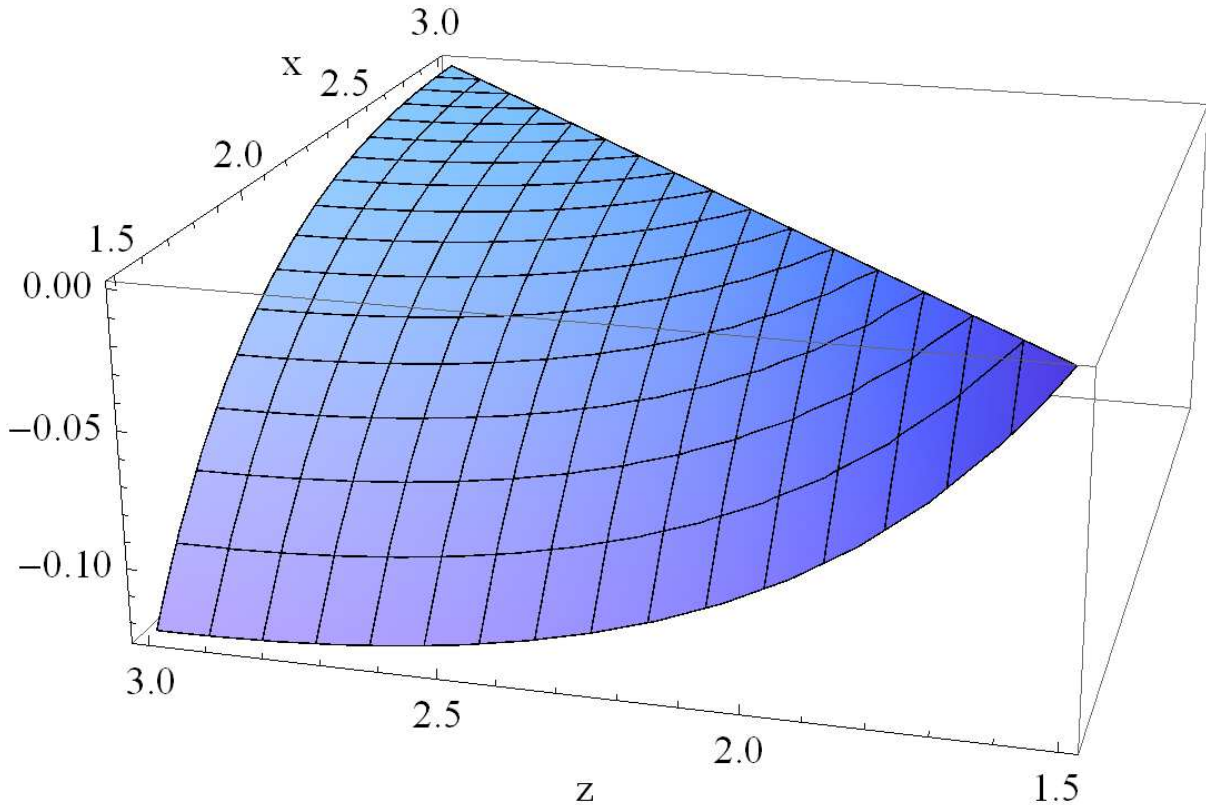}
\end{center}
\caption{The central charge difference $\frac{a_{(A')}+a_{(B)} -a_{(C)}}{N_c^2}$ is always negative in strong gauging of SQCD.}
\label{fig1}
\end{figure}

At this point, we observe that $a_{(A)} + a_{(B)} > a_{(C)}$ as expected from the $a$-theorem, but
we further observe that  $a_{(A')} + a_{(B)}< a_{(C)}$. Therefore, additional gauging of free matter from the non-trivial RG fixed point in SQCD increases the central charge (although the RG flow induced by  strong gauging is a relevant deformation). The statement $a_{(A')} + a_{(B)}< a_{(C)}$ can be shown analytically by using the explicit formula of $a$ from \eqref{aA'} and \eqref{aC} but it is not illuminating at all, so instead we show the numerical plot in figure \ref{fig1}.

We now argue that strong gauging of the additional free chiral superfields from the RG fixed point in SQCD with the  $N_f$ flavors is a ``relevant deformation". By adding extra matter puts the theory off criticality $g = g^* > \tilde{g}^*$ of the gauge coupling constant, where $g^*$ is the original fixed point (with $N_f$ flavors) and $\tilde{g}^*$ is that of after strong gauging (with $N_f+N_f'$ flavors).
As discussed in the previous section, at the original fixed point with $g=g^*$, the RG beta function of the gauge coupling constant was zero, but after strong gauging, it will become positive due to the extra matter contribution. Then the deformation makes the coupling constant decrease toward $\tilde{g}^*$ and we may regard strong gauging as a  relevant deformation.\footnote{We could imagine the possibility that the RG beta function has a second zero so that the deformation is irrelevant. Although theoretically possible, this would lead to a more dramatic RG flow than we know in SQCDs.} Along the RG flow, the wavefunction renormalization factor of newly added matter and original matter would approach each other and eventually become identical so that we end up with the fixed point in SQCD with the $SU(N_f+N_f')^2 \times U(1)$ symmetry.
In the next section, we will present more evidence for the ``relevance" of strong gauging by emulating the RG flow as decoupling of ghost matter.

We should note that the ``added perturbation" induced by strong gauging seems non-renormalizable because if we extrapolate the RG flow toward further UV (rather than IR), then the gauge coupling will typically blow-up (Landau-pole problem) unless we have an unexpected second zero of the RG beta function.
Coincidentally, we cannot apply the argument by Komargodski and Schwimmer  \cite{Komargodski:2011vj}, which proved the $a$-theorem that claims that the central charge $a$ must decrease along the relevant RG flow because just after strong gauging, we are not at the  RG fixed point, and the dilaton effective action does not have the well-defined UV limit.

We can obtain the similar results when IR theory $(A')$ is in the magnetic IR free regime (i.e. $ N_c<N_f < \frac{3}{2} N_c$). In this regime, IR theory $(A')$ is given by the IR free $SU(N_f-N_c)$ gauge theory with $N_f$ (dual) fundamental quark superfields and $N_f^2$ gauge singlet (electric) meson superfields. The central charge may be computed as
\begin{align}
a_{(A')} &=  \frac{N_c^2}{48} \left(9(x-1)^2 + 2(x-1) + x^2 \right) 
\end{align}
by treating all the magnetic degrees of freedom as free fields. We now add the {\it electric} matter as if we are in the electric theories as an example of strong gauging.\footnote{Adding magnetic matter is not interesting because we are in the magnetic IR free regime.} After strong gauging we would end up with either magnetic theory with different gauge group $SU(N_f + N_f'-N_c)$ when $ N_c<N_f + N_f' < \frac{3}{2} N_c$ or non-trivial fixed point in the electric theory when $ \frac{3}{2} N_c<N_f + N_f' < 3 N_c$ (which may be realized as a magnetic theory as well). In both cases, we observe that the central charge satisfies  $a_{(A')} + a_{(B)}< a_{(C)}$. Again, strong gauging increases rather than decreases the central charge $a$.

In order to discuss more non-trivial fixed points, we have also studied  one-adjoint SQCD \cite{Kutasov:2003iy} and two-adjoint SQCD \cite{Intriligator:2003mi} . These theories are more non-trivial in the sense that we need the $a$-maximization method \cite{Intriligator:2003jj} to determine the central charge $a$ at the fixed points.

For one-adjoint SQCD, we take the $SU(N_c)$ gauge group with $N_f$ fundamental flavors in addition to one adjoint chiral matter superfield. We do not introduce any superpotential so that the anomaly free R-symmetry is not unique unlike in the case of SQCD.
We also add $2N_c \times N_f'$   free chiral ``spectator" superfields for the purpose of strong gauging later. By taking the Veneziano limit, we again define
\begin{align}
x &= \frac{N_f}{N_c} \cr
z & = \frac{N_f' + N_f}{N_c}  \ .
\end{align}
In order to compute the central charge before and after gauging, we may use the $a$-maximization technique to determine the superconformal R-charge of the adjoint matter $R_{\mathrm{adj}} $ and the fundamental matter $ R_{\mathrm{fund}}$. According to the $a$-maximization method, at the original fixed point $(A')$,
we maximize the trial $a$-function
\begin{align}
a_{\mathrm{\mathrm{trial}}} = a_0 + \frac{3N_c^2}{32}\left( \left( 3(R_{\mathrm{adj}}-1)^3 - (R_{\mathrm{adj}}-1)\right) + 2x \left(3(R_{\mathrm{fund}}-1)^3 - (R_{\mathrm{fund}}-1) \right) \right) \ ,
\end{align}
with the anomaly free condition 
\begin{align}
x R_{\mathrm{fund}} + R_{\mathrm{adj}} = x \ 
\end{align}
to determine the superconformal R-charge.
Here $a_0 = \frac{3N_c^2}{16}$ is the fixed contribution from the vector multipliet.
In this case, the maximization of $a_{\mathrm{trial}}$ determines the superconformal R-charge of the adjoint multiplet as 
\begin{align}
R_{\mathrm{adj}} = \frac{3x^2 -\sqrt{20x^2-x^4}}{3(-2+x^2)} \ .
\end{align}
The maximum values of $a_{\mathrm{trial}}(x)$ obtained by substituting it back 
gives the central charge $a_{(A')}$ at the original fixed point $(A')$.
Similarly, one may compute the central charge $a_{(C)}$ at the end of strong gauging by replacing $x$ with $z$.

\begin{figure}[tbh]
\begin{center}
\includegraphics[width=\linewidth]{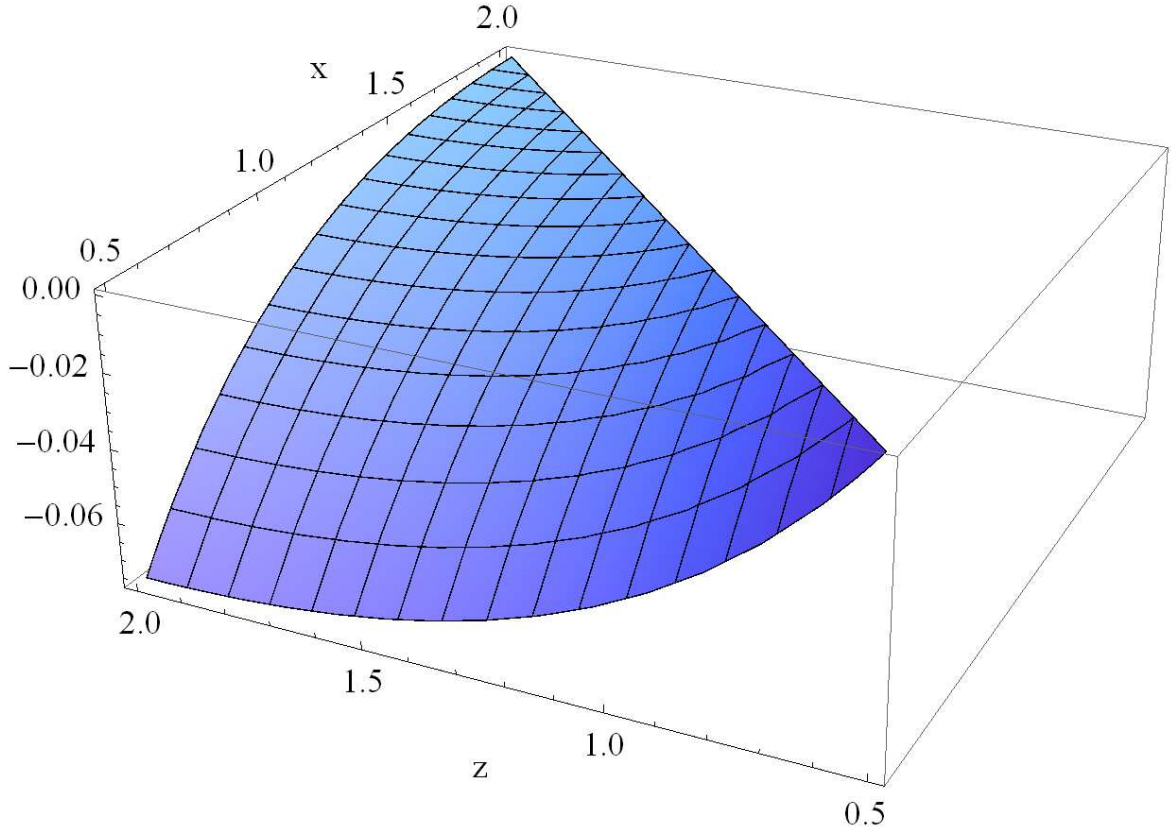}
\end{center}
\caption{The central charge difference $\frac{a_{(A')}+a_{(B)} -a_{(C)}}{N_c^2}$ is always negative in strong gauging of one-adjoint SQCD.}
\label{fig2}
\end{figure}

The resulting central charge difference $a_{(A')} + a_{(B)} - a_{(C)}$ is shown in figure \ref{fig2}. In the figure, we have chosen the range of $x$ such that there is no further accidental $U(1)$ symmetry, but as far as we have studied, the central charge difference is always negative even if we relax the condition.

\begin{figure}[tbh]
\begin{center}
\includegraphics[width=\linewidth]{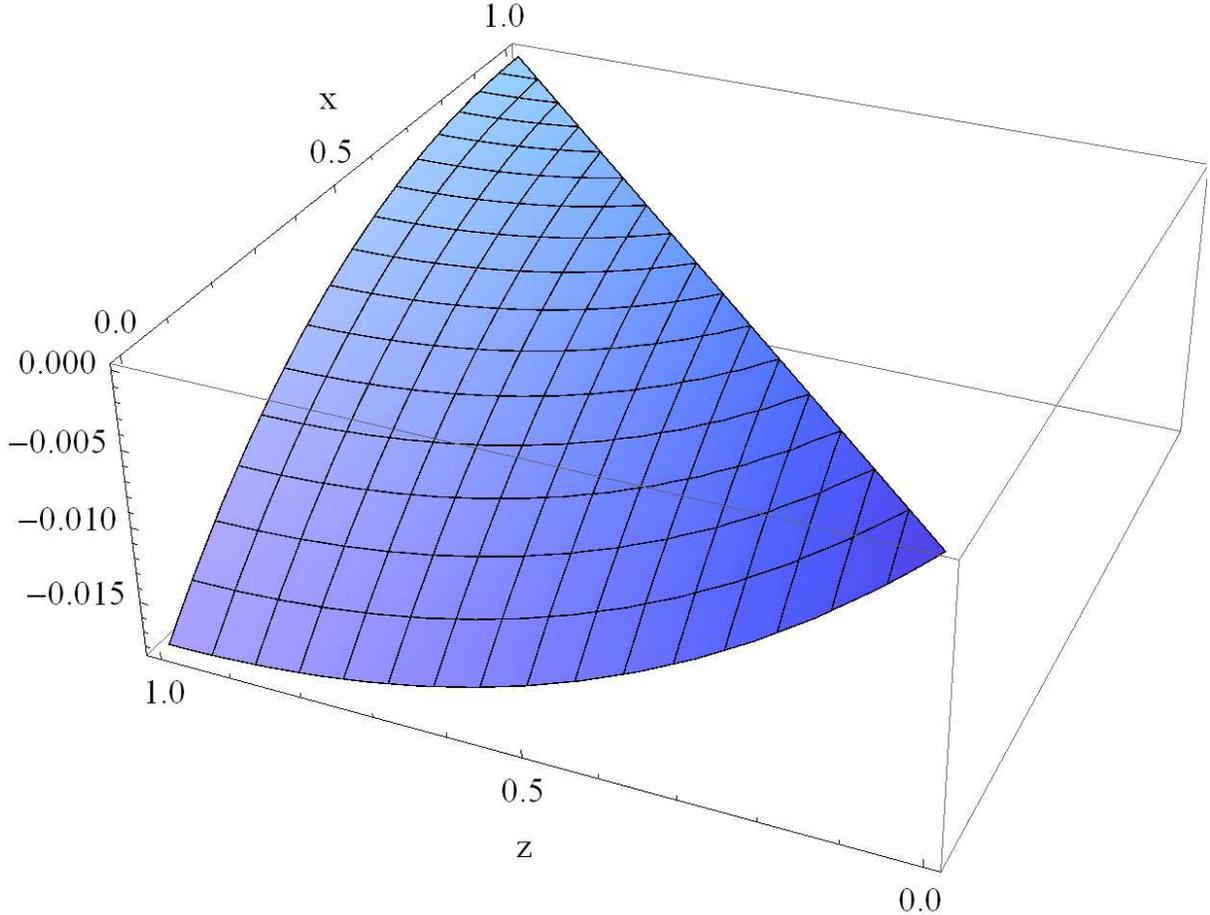}
\end{center}
\caption{The central charge difference $\frac{a_{(A')}+a_{(B)} -a_{(C)}}{N_c^2}$ is always negative in strong gauging of two-adjoint SQCD.}
\label{fig3}
\end{figure}

The similar analysis for the two-adjoint SQCD without superpotential can be found in figure \ref{fig3}. 
The anomaly free condition now becomes
\begin{align}
x(R_{\mathrm{fund}}-1) = 1-2R_{\mathrm{adj}} \ .
\end{align}
By maximizing the trial $a$-function
\begin{align}
a_{\mathrm{trial}} =  a_0 + \frac{3N_c^2}{32}\left( 2\left( 3(R_{\mathrm{adj}}-1)^3 - (R_{\mathrm{adj}}-1)\right) + 2x \left(3(R_{\mathrm{fund}}-1)^3 - (R_{\mathrm{fund}}-1) \right) \right) \ , 
\end{align}
we determine the superconformal R-charge as
\begin{align}
R_{\mathrm{adj}} = \frac{-12 + 3x^2 + \sqrt{26x^2-x^4}}{3(-8+x^2)} \ .
\end{align}
The central charge can be computed by substituting it back to $a_{\mathrm{trial}}$.

In this case, again the central charge increases along the RG flow induced by strong gauging.
We have tried several other examples including the Singlet SQCD \cite{Barnes:2004jj} as well as some quiver gauge theories  \cite{Barnes:2005zn} in which the $a$-maxization technique is heavily used. In every case we have studied without adding any superpotential terms for the newly added matter superfields, the central charge $a$ increases rather than decreases under the RG flow induced by strong gauging.

One may naturally wonder if strong gauging always accompanies an increase of the central charge. While this seems to be the case in most examples of simple strong gauging, probably the question itself is moot without further assumptions about the nature of strong gauging.
The point is that we can always add more interaction terms such as the superpotential coupling between $(A')$ and $(B)$ when we introduce strong gauging. This extra superpotential terms will typically decrease the central charge.\footnote{Due to the $a$-theorem, whatever extra thing we do will cause a decrease of the central charge.}
It is easy to come up with the examples in which the IR central charge will be smaller than the sum of the central charges of $(A')$ and $(B)$ before gauging by combining strong gauging and the introduction of the superpotential. 
We stress that the introduction of the additional superpotential may be sometimes mandatory if we try to keep symmetries such as the flavor symmetry of $SU(N_f+N_f')$ or $\mathcal{N}=2$ supersymmetry. 
In the non-supersymmetric situations, we may have needed to introduce Yukawa-terms or potential terms to reach the IR fixed point from the beginning anyway, so there is no ``canonical" definition of strong gauging. In these situations, there is no surprise that the central charge either decreases or increases depending on the details.
The only point we would like to emphasize in this section is that there is a (large) chance of an increase of the central charge in strong gauging. We cannot claim that  strong gauging always increases the central charge while in many simple cases it does.

\section{UV completion by ghost matter}
As we have already discussed, one problem of gauging extra matter is it is not a well-defined manipulation within the two decoupled CFTs we started with. Note that this not only applies to strong gauging, but also it applies to weak gauging.
In a Lagrangian description, what we do is to add the operator $\int d^d x A^\mu J_\mu + \cdots$ but  clearly $A^\mu$ is not a gauge invariant observable in the original CFT. Moreover, the physical Hilbert space after gauging is quite different from the original Hilbert space of two decoupled CFTs. This together with the fact that strong gauging may not be UV completed renders the argument based on dilaton effective action non-trivial. Indeed, in section 4, we have seen many examples in which the conclusion of the $a$-theorem does not apply in the RG flow induced by  strong gauging.

In this section, we would like to discuss a novel way to emulate the gauging procedure within the conventional picture of the deformed CFT with a dilaton background by using ghost matter. The argument best applies when added matter is vector-like because we will rely on decoupling of ghost matter by introducing its mass term eventually. When the added matter field is chiral, adding its mass term becomes more complicated.
Although we are quite sure that the ghost matter construction we discuss below gives a qualitatively good picture as an emulation of the gauging procedure, the description may not be exact. Nevertheless we will gain some understanding of the mechanism how the central charge $a$ may increase during the induced RG flow.

Our idea is to identify gauging of extra matter at a certain energy scale $M$  as decoupling of  ghost matter at the scale $M$. 
Ghost matter is represented by fields with the wrong spin statistics (similarly to the Pauli-Villars regulator).
In certain situations, we can provide the formal proof that the construction gives the exactly same path integral. For instance, if we have matter Dirac fermions only with gauge interactions, the integration over the Dirac fermions just gives the powers of functional determinant in the remaining path integral over the gauge fields. Thus, we have  the trivial cancellation between the path integral over the ordinary matter field $\psi$, which is anti-commuting, with the  action $S_{\mathrm{Dirac}} = \bar{\psi} D^\mu \gamma_\mu \psi$ and the ghost matter field $\Psi$, which is commuting, with the ghost action $S_{\mathrm{ghost}} = \bar{\Psi} D^\mu \gamma_\mu \Psi$: 
\begin{align}
Z_{N_f} &= \int \mathcal{D} {A_\mu} \mathcal{D}\psi^{N_f} e^{-S_{\mathrm{gauge}} - \sum_{N_f} S_{\mathrm{Dirac}}} \cr
& = \int \mathcal{D} {A_\mu} (\mathrm{Det}D^\mu \gamma_{\mu})^{N_f} e^{-S_{\mathrm{gauge}}} \cr
&= 
\int \mathcal{D} {A_\mu} (\mathrm{Det}D^\mu \gamma_{\mu})^{N_f}  \frac{ \mathrm{Det}(D^\mu\gamma_\mu)}{\mathrm{Det}(D^\mu\gamma_\mu)} e^{-S_{\mathrm{gauge}}} \cr
& =   \int \mathcal{D} {A_\mu} \mathcal{D}\psi^{N_f+1} \mathcal{D}\Psi  e^{-S_{\mathrm{gauge}} - \sum_{N_f+1}S_{\mathrm{Dirac}} -S_{\mathrm{ghost}}} \ .
\end{align}
Adding matter changes the power by one unit but decoupling ghost matter has the same effect. Therefore, we may emulate adding extra matter by decoupling ghost as
\begin{align}
Z_{N_f+1} &= \int \mathcal{D} {A_\mu} \mathcal{D}\psi^{N_f+1} e^{-S_{\mathrm{gauge}} - \sum_{N_f+1}S_{\mathrm{Dirac}}} \cr
&= 
\int \mathcal{D} {A_\mu} (\mathrm{Det}D^\mu \gamma_{\mu})^{N_f+1} e^{-S_{\mathrm{gauge}}} \cr
&= \lim_{M\to \infty} \int \mathcal{D} {A_\mu} \mathrm{Det}(D^\mu \gamma_{\mu})^{N_f} \frac{ \mathrm{Det}(D^\mu\gamma_\mu)}{\mathrm{Det}(D^\mu\gamma_\mu+M)}  e^{-S_{\mathrm{gauge} }} \cr
& = \lim_{M\to \infty} \int \mathcal{D} {A_\mu} \mathcal{D}\psi^{N_f+1} \mathcal{D}\Psi  e^{-S_{\mathrm{gauge}} - \sum_{N_f+1}S_{\mathrm{Dirac}} -S_{\mathrm{ghost + mass}}}  \ ,   
\end{align}
where $S_{\mathrm{ghost + mass}}=  \bar{\Psi} D^\mu \gamma_\mu \Psi + M\bar{\Psi}\Psi$.
If we are only interested in universal features of the IR fixed point, the mass $M$ does not have to be infinite.
While this formal argument relies on the specific form of the interaction of fermions, we propose that the construction is applicable in more generic situations.

To make it more concrete, let us take the example of the IR limit of SQCD again. 
To emulate the IR limit of SQCD before gauging, we prepare the IR limit of $SU(N_c)$ gauge theories with $N_f + N_f'$ fundamental matters $Q$, $\tilde{Q}$ and $N_f'$ fundamental ghost matters $q$, $\tilde{q}$ as theory $(A')$. 
Here the ghost matter is defined by an anti-commuting Lorentz scalar chiral superfield.
The matter superfields have the flavor symmetry of $SU(N_f +N_f'|N_f')^2 \times U(1)$. The fixed point of this system is equivalent to the original fixed point in SQCD with $N_f$ fundamentals. By equivalence, we mean that as long as we assume that the R-symmetry commutes with the supergroup $SU(N_f +N_f'|N_f')^2$, the NSVZ beta function \cite{Novikov:1985rd} is the same and we expect that the singlet sectors under the flavor supergroup would show the same behavior at the IR fixed point.\footnote{In order for this argument to be valid, the supergroup flavor symmetry should not be spontaneously broken. We make this assumption throughout the paper.} 

Now, in order to emulate gauging of extra $N_f'$ flavors by the IR limit of SQCD, we first add the (ordinary) spectator $2N_cN_f'$ chiral superfields with mass $M$. Simultaneously, we also introduce the ghost mass term  in the superpotential $W = m \tilde{q} q$. This triggers the relevant deformations that can be regarded as an emulation of strong gauging of adding extra matter to SQCD because in the far IR limit $\Lambda \ll m, M$, the theory should  flow to the IR limit of SQCD with $N_f + N_f'$ flavor without any ghost matter as theory $(C)$. Note that since $\mathrm{Str}_{SU(N_f +N_f'|N_f')} 1 = \mathrm{Tr}_{SU(N_f)} 1 = N_f$, the central charge $a$ of our emulation before as well as after the decoupling of ghost matter must be same as that of the ordinary SQCD  with different flavor $N_f$ and $N_f+N_f'$ respectively.

We emphasize that in this emulated RG flow, the added interactions (i.e. the ghost mass and the spectator mass) are well-defined renormalizable deformations of the original SCFTs, and it captures the relevant aspects of what happens after  strong gauging in terms of the purely SCFT deformations. At this point, most of our concern about the ``non-renormalizable" or UV non-completed nature of strong gauging we had in section 2 is gone. As for the non-uniqueness of strong gauging, we also had such issues (or rather advantage) here because the equivalent theories with ghost matter are not unique.

Since strong gauging is emulated by adding gauge (or BRST) invariant mass terms, we may apply the dilaton effective field theory argument by Komargodski and Schwimmer. We replace the mass term by the dilaton compensated one $e^{(4-\Delta_i)\tau} O^i$ and study the Wess-Zumino action for the conformal anomaly in the far IR limit. The contribution from the spectator chiral superfields is trivial. It just gives the contribution $\Delta a = \frac{1}{48} 2N_c N_f'$, which decreases the central charge $a$. The positivity of this contribution is guaranteed by the unitarity of the S-matrix for the dilaton scattering as argued in \cite{Komargodski:2011vj}. 
On the other hand, the contribution from decoupling of ghost matter is completely opposite. Since the Hilbert space contains the negative norm sectors, there is no argument for the positivity of the central charge difference. Rather, we would  naturally expect negative contributions to $\Delta a$ because we are decoupling the ghost fields with negative norms. More precisely, the two-point function of the ghost mass term is negative compared with the two-point function of the ordinary matter at the RG fixed point, the conformal perturbation theory \cite{Cardy:1988cwa} predicts that decoupling ghost matter gives the negative (but with the same magnitude due to the supergroup symmetry) contributions to the central charge compared with decoupling ordinary matter at the leading order. Since the supergroup symmetry is broken by the mass term itself, the contributions will differ beyond the leading order.

In order to understand the total central charge difference, we have to compare the gain from decoupling free (ordinary) matter and the loss from decoupling of interacting ghost matter. As we have seen in the previous sections, in many situations that the loss from  decoupling ghost matter is larger because the decoupling typically makes the gauge coupling weaker and less interacting theories typically have larger central charges. 
However, when  ghost matter has the other interactions (e.g. coming from superpotential terms in supersymmetric examples), the decoupling may affect more in the other sectors and the loss of the central charge could be smaller than the gain from decoupling free (ordinary) matter. As we have discussed at the end of section 4, such effects are non-universal and it relies on the details of the ghost matter interaction (and the ordinary matter interaction that we added).\footnote{Some other examples of interacting ghosts include $Sp(N)$ vector models \cite{Anninos:2011ui}\cite{Fei:2015kta}, supermanifold sigma models or their cosets \cite{Witten:2003nn}\cite{Aganagic:2004yh}\cite{Mitev:2011zza}\cite{Quella:2013oda}.}

We finally remark that the introduction of ghost matter may be beautifully realized in ghost D-brane constructions \cite{Okuda:2006fb}. The ghost D-brane is defined as the boundary states with negative coefficients in NS-NS sector as well as R-R sector\footnote{In comparison, the anti D-brane has negative coefficients only in R-R sector, but not in NS-NS sector.} such that the  sum of the ordinary D-brane and ghost D-brane completely cancels in perturbative string amplitudes. Unlike the anti-D-brane it preserves the same supersymmetry as the ordinary D-brane (if placed at an appropriate position in space-time). We can engineer our emulation of decoupling ghost matter by studying webs of D-branes and ghost D-branes similarly to the engineering of SQCD (with additional potentials) from D-brane constructions. 
Various duality arguments on D-branes still hold with ghost D-branes and they would give us  a new method to study our proposed strong gauging through ghost D-branes.

\section{Discussions}

In this paper, we have discussed the implementation of gauging extra matter in strongly coupled gauge theories at criticality. When the added matter fields are vector-like, it is possible to emulate strong gauging by decoupling of ghost matter. This construction when available yields the UV completed definition of strong gauging within a conventional CFT language.

From the perspective of decoupling ghost, the qualitative picture of strong gauging by adding matter is no more different from adding mass term to the ordinary matter, albeit it is ghost now. The problem reduces to what happens after decoupling extra (ghost) matter in a CFT.
When the anomalous dimension of the (ghost) mass term becomes large and the operator dimension of the deformation becomes near marginal, conformal perturbation theory may be used to discuss its fate. This possibility is analogous to the integrable $O_{1,3}$ flow in the Landau-Ginzburg models in $d=2$ dimensions we mentioned in the introduction.  However, in  variants of SQCDs we have studied, the anomalous dimensions of the (ghost) quark superfields happened to be negative, so the supersymmetric decoupling may not be treated by a naive conformal perturbation theory.

Note however that the typicality of the negative anomalous dimensions only apply to the  chiral operators that could be added in the superpotential. In contrast, the non-supersymmetric mass terms such as the Konishi operator always have positive anomalous dimensions. 
Thus, there is a possibility that adding or decoupling of (ghost) matter in a non-supersymmetric situation can be treated by a conformal perturbation theory.

Within perturbation theory, we believe that the use of ghost matter does not cause any serious drawback. In the supersymmetric situations, the perturbative computations are sometimes non-perturbatively exact as demonstrated in localization of the path integral. However, it is possible that ghost matter may cause the non-perturbative deviation from what we expect in the emulation of strong gauging. This happens, for instance, when the supergroup flavor symmetry is spontaneously broken. 
It would be important to sharpen the conditions under which the emulation would give the incorrect result. While  it is a logically independent problem, it would be interesting to see what happens in actual strong gauging in such situations.  There has been a study on the ghost matrix model in relation to holography \cite{Vafa:2014iua} and the non-perturbative insight there may be useful.

\section*{Acknowledgements}
The author would like to thank Slava Rychkov for correspondence and discussions. He also would like to express his sincere gratitude to the Weizmann Institute of Science for the hospitality during my stay.
This work is supported by a Sherman Fairchild Senior Research Fellowship at the California Institute of Technology  and DOE grant number DE-SC0011632.

\end{document}